\documentstyle[12pt,a4,epsf,cite]{article}
\def\refjl#1#2#3#4#5#6{\bibitem{#1} #2, {\it #3} {\bf #4} (#5) #6.}
\def\refbk#1#2#3#4{\bibitem{#1} #2, {\it #3}, #4.}
\def\etal{{\it et al}}
\def\NP{Nucl. Phys.}
\def\NPPS{Nucl. Phys. B (Proc. Suppl.)}
\def\PL{Phys. Lett.}
\def\PRL{Phys. Rev. Lett.}
\def\PR{Phys. Rev.}

\def\ZP{Z. Phys.}
\def\MPL{Mod. Phys. Lett.}

\def\APNY{Ann. Phys., NY}
\def\NC{Nuovo Cimento}

\newcommand{\eqn}[1]{(\ref{#1})}
\newcommand{\be}{\begin{equation}}
\newcommand{\ee}{\end{equation}}
\newcommand{\no}{\nonumber}
\newcommand{\bel}[1]{\be\label{#1}}
\newcommand{\ba}{\begin{array}{c}}
\newcommand{\bat}{\begin{array}{cc}}
\newcommand{\ea}{\end{array}}
\newcommand{\beqn}{\begin{eqnarray}}
\newcommand{\eeqn}{\end{eqnarray}}

\newcommand{\bi}{\begin{itemize}}
\newcommand{\ei}{\end{itemize}}
\newcommand{\rms}{\rm\scriptsize}

\renewcommand{\thefootnote}{\fnsymbol{footnote}}
\begin{document}
\begin{titlepage}
\begin{flushright}
{FTUV/97-47}\\
{IFIC/97-63}\\
{UG-FT-77/97}\\
{hep-ph/9804462}\\
April 1998  \\
\end{flushright}
\vspace{2cm}
\begin{center}
{\large\bf PERTURBATIVE QUARK MASS CORRECTIONS \\
 TO THE TAU HADRONIC WIDTH}\\
\vfill
{\bf Antonio Pich$^a$ and Joaquim Prades$^b$}\\[0.5cm]
$^a$ Departament de F\'{\i}sica Te\`orica,
 IFIC, Universitat de Val\`encia --- CSIC \\
 Dr. Moliner 50,  E-46100 Burjassot (Val\`encia), Spain.\\[0.5cm]
$^b$ Departamento de F\'{\i}sica Te\'orica y del Cosmos,
Universidad de Granada, Campus de Fuente Nueva, E-18002 Granada, Spain.
\\[0.5cm]
\end{center}
\vfill
\begin{abstract}
The perturbative quark--mass corrections to the $\tau$ hadronic width
are analysed to $O(\alpha_s^3 m_q^2)$, using the presently available
theoretical information. The behaviour of the perturbative series is
investigated in order to assess the associated uncertainties.
The implications for the determination of the strange quark
mass from $\tau$ decay data are discussed.
\end{abstract}
\vspace*{1cm}
\vfill
\end{titlepage}

\setcounter{footnote}{0}
\renewcommand{\thefootnote}{\alph{footnote}}


\vspace{0.5cm}
\noindent {\bf 1. \, Introduction}
\vspace{0.5cm}

The inclusive character of the total $\tau$ hadronic width renders possible
an accurate calculation of the ratio
\cite{BR:88,NP:88,BNP:92,LDP:92a,PI:97}
\be\label{eq:r_tau_def}
  R_\tau\equiv {\Gamma \left[\tau^-\rightarrow\nu_\tau
  \,\mbox{\rm hadrons}\, (\gamma)\right]\over\Gamma \left[\tau^-\rightarrow
  \nu_\tau e^- {\bar \nu}_e\, (\gamma)\right] } \; ,
\ee
using standard field theory methods. The result turns out to be very
sensitive to the value of $\alpha_s(M_\tau^2)$. Moreover, the uncertainties
in the theoretical calculation are quite small and dominated by the
perturbative errors. This has been used to perform a very precise
determination of the QCD coupling at low energies \cite{PI:97}.

Quark masses play a rather minor r\^ole in $R_\tau$. Owing to the tiny
values of $m_u$ and $m_d$, their associated corrections are very small
\cite{BNP:92} ($\sim - 0.1\%$).
The strange quark contribution to the total $\tau$ hadronic
width is suppressed by the Cabibbo factor $|V_{us}|^2$,
which puts the induced $m_s$ correction also at the
per cent level. However, if one analyses separately the  semi-inclusive
decay width of the $\tau$ into Cabibbo--suppressed modes (i.e. final states
with an odd number of kaons), the relatively large value of $m_s$ induces
an important effect of a size similar to the massless perturbative
correction and of opposite sign \cite{BNP:92}.
The corresponding $R_{\tau,S}$ prediction is then very sensitive to
the strange quark mass and could be used to extract information on this
important, and nowadays controversial, parameter.
A very preliminary value of $m_s$, extracted from the ALEPH
$\tau$ decay data, has been already presented in recent workshops
\cite{Chen:98,DA:97}.

The determinations of light quark masses are usually obtained from
analyses of the divergences of the vector and axial--vector current
two--point function correlators or related observables
\cite{BE:81,DdeR:87,JM:95,BPdeR:95,JA:98,PR:98}.
These correlators are proportional to quark masses and, therefore, are
very sensitive to their numerical values. Unfortunately, one needs
phenomenological information on the associated scalar and
pseudo-scalar spectral functions, which are not well known at
present.
The obvious advantage of a possible
determination of $m_s$ analysing quark mass effects in $\tau$ decays
is that the experimental
error can be systematically reduced in foreseen facilities like tau--charm
or B factories.
There is then some hope to achieve a precise determination of
$m_s$ from such analyses.

Recently the $O(\alpha_s^3)$ corrections to the $J=0$
quark correlators have been calculated \cite{CH:97},
and have been found to be rather large.
The influence of these $O(\alpha_s^3)$ corrections on the
determination of quark masses and the uncertainties
coming from the truncation of the QCD perturbative series depend
very much on the observable.
One can see for instance that the QCD perturbative series
behaves geometrically to $O(\alpha_s^3)$
for the divergence of pseudo-scalar (scalar)
currents if resummed perturbatively in terms of $\alpha_s(s)$
\cite{JA:98,PR:98}.
This convergence improves \cite{PR:98} using other resummations like the
Principle of Minimal Sensitivity (PMS) \cite{ST:81}
or the one advocated in Ref.~\cite{LDP:92a}.

The hadronic $\tau$ decay width has also a $J=0$ contribution,
which, as we shall see, behaves rather badly.  However,
the largest quark--mass correction originates in a
piece of the left--handed current correlation function,
involving the $J=0+1$ combination, which shows a much
better perturbative convergence.

The purpose of this paper is to study the perturbative behaviour of the
corrections to $R_\tau$ which are proportional to $m_q^2$, in order to
assess the associated uncertainties. These are the leading theoretical
uncertainties in the $m_s$ determination.
The $O(\alpha_s m_q^2)$ contributions were
already studied in Ref.~\cite{BNP:92}. In Ref.~\cite{CK:93} the
contributions of $O(\alpha_s^2 m_q^2)$ to the relevant correlators were
worked out. More recently, some partial information on $O(\alpha_s^3 m_q^2)$
corrections has become available \cite{CH:97,CK:97}.
A much detailed analysis of all contributions
up to dimension four will be presented  elsewhere.

\vspace{0.5cm}
\noindent {\bf 2. \, Theoretical Framework}
\vspace{0.5cm}

The theoretical analysis of $R_\tau$ involves
the two--point correlation functions for
the vector $\, V^{\mu}_{ij} = \bar{\psi}_j \gamma^{\mu} \psi_i \, $
and axial--vector
$\, A^{\mu}_{ij} = \bar{\psi}_j \gamma^{\mu} \gamma_5 \psi_i \,$
colour--singlet quark currents ($i,j=u,d,s$):
\beqn\label{eq:pi_v}
\Pi^{\mu \nu}_{ij,V}(q) &\!\!\! \equiv &\!\!\!
 i \int d^4x \, e^{iqx}
\langle 0|T(V^{\mu}_{ij}(x) V^{\nu}_{ij}(0)^\dagger)|0\rangle  ,
\\
\label{eq:pi_a}
\Pi^{\mu \nu}_{ij,A}(q) &\!\!\! \equiv &\!\!\!
 i \int d^4x \, e^{iqx}
\langle 0|T(A^{\mu}_{ij}(x) A^{\nu}_{ij}(0)^\dagger)|0\rangle  .
\eeqn
They have the Lorentz decompositions
\bel{eq:lorentz}
\Pi^{\mu \nu}_{ij,V/A}(q)  =
  (-g^{\mu\nu} q^2 + q^{\mu} q^{\nu}) \, \Pi_{ij,V/A}^{(1)}(q^2)
  +   q^{\mu} q^{\nu} \, \Pi_{ij,V/A}^{(0)}(q^2) ,
\ee
where the superscript $(J)$
in the transverse and longitudinal components denotes the corresponding
angular momentum $J=1$ (T) and $J=0$ (L) in the hadronic rest frame.

The imaginary parts of the two--point functions
$\, \Pi^{(J)}_{ij,V/A}(q^2) \, $
are proportional to the spectral functions for hadrons with the corresponding
quantum numbers.  The semi-hadronic decay rate of the $\tau$
can be written as an integral of these spectral functions
over the invariant mass $s$ of the final--state hadrons:
\bel{eq:spectral}
R_\tau   =
12 \pi \int^{M_\tau^2}_0 {ds \over M_\tau^2 } \,
 \left(1-{s \over M_\tau^2}\right)^2
\biggl[ \left(1 + 2 {s \over M_\tau^2}\right)
 \mbox{\rm Im} \Pi^{(1)}(s)
 + \mbox{\rm Im} \Pi^{(0)}(s) \biggr]  .
\ee
The appropriate combinations of correlators are
\bel{eq:pi}
\Pi^{(J)}(s)  \equiv
  |V_{ud}|^2 \, \left( \Pi^{(J)}_{ud,V}(s) + \Pi^{(J)}_{ud,A}(s) \right)
+ |V_{us}|^2 \, \left( \Pi^{(J)}_{us,V}(s) + \Pi^{(J)}_{us,A}(s) \right).
\ee

We can decompose the predictions for $R_\tau$ into contributions
associated with specific quark currents:
\be\label{eq:r_tau_v,a,s}
 R_\tau \, = \, R_{\tau,V} + R_{\tau,A} + R_{\tau,S}\, .
\ee
$R_{\tau,V}$ and $R_{\tau,A}$ correspond to the contributions from the first
two terms in Eq.~\eqn{eq:pi}, while
$R_{\tau,S}$ contains the remaining Cabibbo--suppressed contributions.

Exploiting the analytic properties of the correlators $\Pi^{(J)}(s)$,
Eq.~(\ref{eq:spectral})
can be expressed as a contour integral in the complex $s$ plane
running counter--clockwise around the circle $|s|=M_\tau^2$:
\bel{eq:circle}
R_\tau  =
-  \pi i \oint_{|s|=M_\tau^2} {ds \over s} \,
 \left(1 - {s \over M_\tau^2}\right)^3
 \left\{ 3 \left(1 + {s \over M_\tau^2}\right) D^{L+T}(s)
        + 4 \, D^{L}(s) \right\} .
\ee
We have used integration by parts to rewrite $R_\tau$ in terms of
the logarithmic derivative of the relevant correlators,
\be
D^{L+T}(s) \equiv - s\, {d\over d s} \,\left[\Pi^{(0+1)}(s) \right]\, ,
\qquad
D^{L}(s) \equiv   {s\over M_\tau^2}\, {d\over d s}\,
\left[s \,\Pi^{(0)}(s) \right]\, ,
\ee
which satisfy homogeneous renormalization group equations.

Using the Operator Product Expansion to organise the perturbative and
non-perturbative contributions to the correlators into a systematic
expansion \cite{SVZ:79} in powers of $1/s$, the total ratio $R_\tau$
can be expressed as an expansion in powers of $1/M_\tau^2$, with coefficients
that depend only logarithmically on $M_\tau$ \cite{BNP:92}:
\be  \label{eq:r_total}
R_{\tau}  =  3 \left( |V_{ud}|^2 + |V_{us}|^2 \right)
S_{EW} \biggl\{ 1 + \delta_{EW}' + \delta^{(0)}
 + \sum_{D=2,4,...}
         \left( \cos^2 \theta_C \delta^{(D)}_{ud}
         + \sin^2 \theta_C \delta^{(D)}_{us} \right) \biggr\} ,
\ee
where
$\delta^{(D)}_{ij} = (\delta^{(D)}_{ij,V} + \delta^{(D)}_{ij,A})/2$
is the average of the vector and axial--vector corrections of dimension
$D$, $S_{EW}$ and $\delta_{EW}'$ contain the known \cite{MS:88,BL:90}
electroweak corrections, and
$\sin^2\theta_C\equiv |V_{us}|^2/(|V_{ud}|^2 + |V_{us}|^2)$.

The dimension--zero contribution    
 is the purely perturbative correction
neglecting quark masses, which, owing to chiral symmetry, is identical for
the vector and axial--vector correlators. It is fully generated by
the Adler function $D^{L+T}(s)$, because $D^{L}(s)$ vanishes in the
chiral limit. The correction
$\delta^{(0)}$ has been investigated in great detail in
Ref.~\cite{LDP:92a}. We will follow a similar procedure to analyse
the perturbative quark--mass corrections of dimension two.

\vspace{0.5cm}
\noindent {\bf 3. \, Dimension--Two Corrections}
\vspace{0.5cm}

For the sake of simplicity, let us take here $m_u = m_d = 0$.
The arguments we shall put forward don't depend on it.
In this limit, the vector
and axial--vector correlators get the same quark--mass  
corrections, i.e.
$D^J_{us,A}(s) = D^J_{us,V}(s) \equiv D^J_{us}(s)\ $
($J = L+T, L$).
The dimension--two contributions can be written in the form:
\beqn\label{eq:AdlerD2(0+1)}
D^{L+T}_{us}(s)\big|_{D=2} &\!\!\! = &\!\!\!
  \phantom{-}  {3\over 4 \pi^2}\, {m_s^2(-\xi^2 s)\over s}
  \;\sum_{n=0} \tilde{d}^{L+T}_n(\xi)\, a^n(-\xi^2 s) \, ,
\\ \label{eq:AdlerD2(0)}
D^{L}_{us}(s)\big|_{D=2} &\!\!\! = &\!\!\!
  - {3\over 8 \pi^2} \, {m_s^2(-\xi^2 s)\over M_\tau^2}
  \;\sum_{n=0} \tilde{d}^L_n(\xi)\, a^n(-\xi^2 s) \, ,
\eeqn
where $a = \alpha_s/\pi$, $\xi$ is an arbitrary scale factor (of order unity)
and the coefficients $\tilde{d}^J_n(\xi)\ $  are constrained
by the homogeneous renormalization group equations satisfied by
the corresponding functions $D^J_{us}(s)$:
\bel{eq:rge_d}
\xi \, {d\over d \xi} \,\tilde{d}^J_n(\xi) = \sum_{k=1}^n
\left[ 2 \gamma_k - (n-k) \beta_k\right] \, \tilde{d}^J_{n-k}(\xi) \, ,
\ee
for $n\geq1$ and
\bel{eq:rge_d0}
{d\over d \xi} \,\tilde{d}^J_0(\xi) = 0
\ee
i.e.
\beqn\label{eq:coef}
\tilde{d}^J_0(\xi) &\!\!\! = &\!\!\!  d^J_0 \, ,
\no\\
\tilde{d}^J_1(\xi) &\!\!\! = &\!\!\!
d^J_1 + 2 \gamma_1  d^J_0 \log{\xi} \, ,
\no\\
\tilde{d}^J_2(\xi) &\!\!\! = &\!\!\!
 d^J_2 + \left[ 2 \gamma_2 d^J_0 + (2 \gamma_1 - \beta_1) d^J_1 \right]
\log{\xi}  + \gamma_1 (2 \gamma_1 - \beta_1)  d^J_0 \log^2{\xi} \, ,
\no \\
\tilde{d}^J_3(\xi) &\!\!\! = &\!\!\!
 d^J_3 + \left[ 2 \gamma_3 d^J_0 + (2 \gamma_2 - \beta_2) d^J_1 +
            2 (\gamma_1 - \beta_1) d^J_2 \right] \log{\xi}
\\ &&\!\!\! \mbox{}
 + \left[
   \left(-\gamma_1 \beta_2 + 2 \gamma_2 (2 \gamma_1 - \beta_1)\right) d^J_0  +
     (\gamma_1 - \beta_1)  (2 \gamma_1 - \beta_1) d^J_1 \right]  \log^2{\xi}
\no\\ &&\!\!\! \mbox{}
 + {2\over 3} \gamma_1 (\gamma_1 - \beta_1)  (2 \gamma_1 - \beta_1) d^J_0
  \log^3{\xi} \, ,
\no\\
\tilde{d}^J_4(\xi) &\!\!\! = &\!\!\! d^J_4 +
\cdots \no
\eeqn
The factors $\beta_k$ and $\gamma_k$ are the expansion coefficients of the
QCD $\beta$ and $\gamma$ functions,
\bel{eq:beta}
\mu\, {d a\over d \mu} = \beta(a) \, a \, , \qquad\qquad
\beta(a) = \sum_{k=1} \beta_k \, a^k \, ,
\ee
\bel{eq:gamma}
\mu \, {d m\over d \mu} = - \gamma(a)\, m \, , \qquad\qquad
\gamma(a) = \sum_{k=1} \gamma_k\, a^k \, ,
\ee
which are known to four loops
\cite{RVL:97,CKS:97,CH:97b}.    
 The coefficients
$d^J_n\equiv\tilde{d}^J_n(1)$
are only known to order $\alpha_s^2$ for $J=L+T$
and $\alpha_s^3$ for $J=L$
\cite{BE:81,CH:97,CK:97,S89,GKLS90,GKLS91,GKL86,BW81}.

For three flavours and in the $\overline{\mbox{MS}}$ scheme, one has:
$$
\beta_1 \,  = \, -{9\over 2} \, ; \qquad\qquad \beta_2  \, = \, -8 \, ;
\qquad\qquad
 \beta_3  \, = \, -{3863\over 192} \, ;
$$
\bel{eq:beta_k}
\beta_4 \, = \,  - {140599 \over 2304} - {445 \over 16}\, \zeta_3
\, \approx\, - 94.456 \, 079 \; ;
\ee
$$
\gamma_1  \, = \, 2 \; ; \qquad\qquad \gamma_2  \, = \, {91 \over 12} \; ;
\qquad\qquad
\gamma_3  \, = \, {8885\over 288} - 5 \,\zeta_3  \, \approx \, 24.840\, 410 \;
;
$$
\bel{eq:gamma_k}
\gamma_4  \, = \, {2977517 \over 20736} - {9295 \over 216} \,\zeta_3
   + {135 \over 8} \,\zeta_4- {125 \over 6} \,\zeta_5
   \, \approx \, 88.525\, 817\; ;
\ee
$$   
d_0^{L+T}  =  1 \; ;\qquad
d^{L+T}_1  =  {13\over 3}\; ; \qquad
d^{L+T}_2  =  {21541 \over 432} + {323 \over 54}\,  \zeta_3
    - {520\over 27} \,\zeta_5  \, \approx \, 37.083\, 047\, ;
$$
\bel{eq:d^L_n}
d_0^L  =  1 \; ;\qquad\quad
d^L_1  =  {17\over 3} \; ; \qquad\quad
d^L_2  =  {9631 \over 144} - {35 \over 2}  \,\zeta_3
 \, \approx \, 45.845\, 949\; ;
\ee
$$
d^L_3  =  {4748953 \over 5184} - {91519 \over 216}  \,\zeta_3
    - {5\over 2} \,\zeta_4 + {715\over 12} \,\zeta_5  \, \approx \, 465.846\,
304\, .
$$
Notice the rather bad perturbative behaviour of the $D=2$ corrections
to the correlation functions $D^J_{us}$.
Remember that $a(M_\tau^2)\simeq0.11$.

Inserting the expansions \eqn{eq:AdlerD2(0+1)} and \eqn{eq:AdlerD2(0)}
in Eq.~\eqn{eq:circle}, the $D=2$ corrections to $R_{\tau,S}$ can be
expressed as
\bel{eq:d2form}
\delta^{(2)}_{us} = - 8 \,{m_s^2(M_\tau^2)\over M_\tau^2}\;\Delta[a(M_\tau^2)]
\, , \qquad\qquad
\Delta[a] \equiv {1\over 4}
  \left\{ 3\, \Delta^{L+T}[a] + \Delta^L[a]\right\} \, ,
\ee
where
\bel{eq:Delta}
\Delta^J[a(M_\tau^2)] \, = \, \sum_{n=0}
  \tilde{d}^J_n(\xi) \, B_J^{(n)}(a_\xi) \, ,
\ee
and the contour integrations are contained in the functions
\beqn\label{eq:B_T}
B_{L+T}^{(n)}(a_\xi) &\!\!\!\equiv &\!\!\!
 {-1\over 4 \pi  i} \,\oint_{|x|=1}
  {d x\over x^2}\, (1+x) \, (1 -x)^3
 \left( {m(-\xi^2 M_\tau^2 x)\over m(M_\tau^2)}\right)^2
 a^n(-\xi^2 M_\tau^2 x)  , \quad
\\ \label{eq:B_L}
B_L^{(n)}(a_\xi) &\!\!\!\equiv &\!\!\!
{1\over 2 \pi  i} \,\oint_{|x|=1}
 \, {d x\over x} \, (1 -x)^3 \,
 \left( {m(-\xi^2 M_\tau^2 x)\over m(M_\tau^2)}\right)^2 \,
 a^n(-\xi^2 M_\tau^2 x) \, .
\eeqn
Since the quark mass ratio is flavour independent,
the integrals $B_J^{(n)}(a_\xi)$ regulate also the small
corrections proportional to $m_u$ and $m_d$, which we are neglecting.
These functions depend only on  $\beta_i$, $\gamma_j$,
$a_\xi \equiv \alpha_s(\xi^2 M_\tau^2)/\pi$  and $\log{\xi}$. Moreover,
they satisfy the following homogeneous renormalization group equations:
\bel{eq:B_rge}
\xi \, {d\over d \xi}\, B_J^{(n)}(a_\xi)
 = \sum_{k=1}
\left( n \beta_k - 2 \gamma_k\right) \, B_J^{(n+k)}(a_\xi)
 \, .
\ee

\vspace{0.5cm}
\noindent {\bf 4. \, Perturbative $a_\xi$ Expansion}
\vspace{0.5cm}

The usual perturbative approach expands the $B_J^{(n)}(a_\xi)$
 functions in powers of $a_\xi$. This gives,
\beqn\label{eq:Bexp}
\lefteqn{B_J^{(0)}(a_\xi) =
 1 - \gamma_1 \left[ 2\log{\xi} + H_1^J \right] a_\xi}
&& \no\\ &&  \mbox{}
-\Biggl[
  2 \gamma_2 \log{\xi} -\gamma_1 (2\gamma_1 +\beta_1) \log^2{\xi}
  + \left( \gamma_2 - 2 \gamma_1^2 \log{\xi} \right) H^J_1
  \Biggr. \no\\ && \Biggl. \quad\;\mbox{}
  - {\gamma_1\over 2} \left( \gamma_1 - {\beta_1\over 2}\right) H^J_2
\Biggr] a_\xi^2
\no\\ &&  \mbox{}
- \Biggl[
      2\gamma_3 \log{\xi}
  - \left( 2 \gamma_2 (2\gamma_1+\beta_1) +\gamma_1\beta_2\right)\log^2{\xi}
  + {2\over 3}\gamma_1 (\gamma_1+\beta_1) (2 \gamma_1+\beta_1) \log^3{\xi}
  \Biggr. \no\\ && \Biggl. \quad\;\mbox{}
  + \left(  \gamma_3 - 4\gamma_1\gamma_2 \log{\xi}
     + \gamma_1^2 (2\gamma_1 + \beta_1) \log^2{\xi}
    \right) H^J_1
  \Biggr. \no\\ && \Biggl. \quad\;\mbox{}
  + \left( {1\over 2} \beta_1 \gamma_2 + {1\over 4} \beta_2\gamma_1
     -\gamma_1\gamma_2
  + \gamma_1^2 \left(\gamma_1 - {\beta_1\over 2}\right) \log{\xi}
    \right) H^J_2
  \Biggr. \no\\ && \Biggl. \quad\;\mbox{}
+ {\gamma_1\over 12} (2\gamma_1-\beta_1) (\gamma_1-\beta_1) H^J_3
\Biggr] a_\xi^3
\, + \, \cdots
\\ \no\\
\lefteqn{B_J^{(1)}(a_\xi) =
 a_\xi \ - \left[2 \gamma_1 \log{\xi}
+ \left( \gamma_1 - {\beta_1\over2}\right) H_1^J \right] a_\xi^2}
&& \no\\ &&  \mbox{}
-\Biggl[
  2\gamma_2 \log(\xi) - \gamma_1 (2\gamma_1 + \beta_1) \log^2{\xi}
 + \left( \gamma_2 - {\beta_2\over 2} -
       \gamma_1 (2\gamma_1 - \beta_1) \log{\xi} \right) H^J_1
\Biggr. \no\\ &&  \Biggl.\quad\;\mbox{}
 - {1\over 2}\left( \gamma_1 - {\beta_1\over2}\right)
   \left(\gamma_1-\beta_1\right) H^J_2
\Biggr] a_\xi^3
\, + \,\cdots
\\ \no\\
\lefteqn{B_J^{(2)}(a_\xi) =
 a_\xi^2 - \left[2 \gamma_1 \log{\xi} +
 \left( \gamma_1 - \beta_1 \right)
 H^J_1 \right] a_\xi^3 + \cdots } &&
\\ \no\\
\lefteqn{B_J^{(3)}(a_\xi) =
 a_\xi^3 - \cdots } &&
\eeqn
where
\bel{eq:H^T_integ}
H^{L+T}_n \equiv {-1\over 4\pi i}\,\oint_{|x|=1}
  {d x\over x^2}\, (1+x) \, (1 -x)^3 \log^n{(-x)} \; ,
\ee
\bel{eq:H^L_integ}
H^L_n \equiv {1\over 2\pi i}\,\oint_{|x|=1}
  {d x\over x}\,  (1 -x)^3 \log^n{(-x)} \; .
\ee
To $O(\alpha_s^4)$, the needed integrals are
$$
H^{L+T}_0  =  1 \; , \qquad\qquad H^{L+T}_1 = {1\over 6} \; , \qquad\qquad
H^{L+T}_2 = {25\over 18} - {\pi^2\over 3} \; ,
$$
\bel{eq:H^T_n}
H^{L+T}_3 = {85\over 36} - {\pi^2\over 6} \; , \qquad\qquad
H^{L+T}_4  = {721\over 54} - {25\over 9}\,\pi^2 + {\pi^4\over 5}\; ,
\ee
$$
H^L_0  = 1 \; , \qquad\qquad H^L_1 = - {11\over 6} \; , \qquad\qquad
H^L_2 = {85\over 18} - {\pi^2\over 3} \; ,
$$
\bel{eq:H^L_n}
H^L_3 = - {575\over 36} + {11\over 6}\, \pi^2 \; , \qquad\qquad
H^L_4  = {3661\over 54} - {85\over 9}\,\pi^2 + {\pi^4\over 5}\; .
\ee

The perturbative expansions $\Delta^J[a]$ then take the form
\bel{eq:Delta_exp}
\Delta^J[a(M_\tau^2)] \, = \, \sum_{n=0}
\left[ \tilde{d}^J_n(\xi) + \tilde{h}^J_n(\xi)\right] \, a_\xi^n \; ,
\ee
where the coefficients $\tilde{h}^J_n(\xi)$ depend on $\tilde{d}^J_{m<n}(\xi)$,
$\beta_{m<n}$ and $\gamma_{m\leq n}$; thus, they are known up to
$O(a^3)$ and $O(a^4)$ for $J=L+T$ and $J=L$ respectively.
For $\xi=1$, one has [$h^J_n\equiv\tilde{h}^J_n(1)$]:
$$
h^{L+T}_0  =  0 \; ; \qquad\qquad h^{L+T}_1 = -{1\over 3} \; ; \qquad\qquad
h^{L+T}_2 = {113\over 72} - {17\over 12} \,\pi^2 \approx -12.412\, 495\; ;
$$
\bel{eq:h^T}
h^{L+T}_3  =  {114517\over 2592} - {4391\over 144}\, \pi^2
-{3659\over 648}\, \zeta_3 + {1690\over 81}\,\zeta_5 \approx -241.926\, 329\; ;
\ee
\beqn
h^{L+T}_4 &\!\!\! = &\!\!\! {26864009\over 13824} - {3110783\over 5184}\, \pi^2
 - {1073\over 1920}\, \pi^4 + {3051761\over 15552}\, \zeta_3
 - {123745\over 2592}\, \pi^2\zeta_3
\no\\ &&\!\!\! \mbox{}
 - {171845\over 243}\,\zeta_5
 + {29575\over 162}\, \pi^2\zeta_5 - {35\over 24} \, d_3^{L+T}
 \approx -3\, 229.101\, 787  - {35\over 24} \, d_3^{L+T}\; ;
\no\\ \no
\eeqn
$$
h^L_0  =  0 \; ; \qquad\qquad h^L_1 = {11\over 3} \; ; \qquad\qquad
h^L_2 = {625\over 8} - {17\over 12} \,\pi^2 \approx 64.143\, 060\; ;
$$
\bel{eq:h^L}
h^L_3  =  {1435691\over 864} - {7927\over 144}\, \pi^2
 -{5225\over 24}\, \zeta_3 \approx 856.673\, 579\; ;
\ee
\beqn
h^L_4 &\!\!\! = &\!\!\! {693706385\over 20736} - {2295071\over 1728}\, \pi^2
 - {10877\over 17280}\, \pi^4 - {1429525\over 144}\, \zeta_3
 + {5595\over 32}\, \pi^2\zeta_3  + {264275\over 288}\,\zeta_5
\no\\ &&\!\!\! \mbox{}
 \approx 11\, 377.111\, 254\; .
\no
\eeqn

 The contour integration generates rather large numerical factors, which show
an opposite behaviour for the transverse and longitudinal pieces. In the
$\Delta^{L+T}$ expansion the $h_n^{L+T}$
 contributions cancel to some extent with the
original correlation--function coefficients $d^{L+T}_n$,
\beqn\label{eq:Delta^T_exp}
\lefteqn{\Delta^{L+T}[a(M_\tau^2)]  =1  + 4 \; a(M_\tau^2) +
  \left( {22219\over 432} - {17\over 12}\,\pi^2 + {323\over 54}\,\zeta_3
  - {520\over 27}\,\zeta_5 \right) \, a(M_\tau^2)^2}
\no\\ &&\mbox{}   +
\left( d^{L+T}_3+{114517\over 2592} - {4391\over 144}\,\pi^2
  - {3659\over 648}\,\zeta_3
+ {1690\over 81}\,\zeta_5 \right) \, a(M_\tau^2)^3
+ \cdots   \no\\  && =
 1 + 4 \; a(M_\tau^2) + 24.671\; a(M_\tau^2)^2 +
\left( d^{L+T}_3 -  241.926 \right) \, a(M_\tau^2)^3 + \cdots
\eeqn
 However, both $d^L_n$
and $h^L_n$ contributions are large and positive, which gives rise to a badly
behaved expansion for $\Delta^L$:
\beqn\label{eq:Delta^L_exp}
\lefteqn{\Delta^L[a(M_\tau^2)]  = 1 + {28\over 3} \; a(M_\tau^2) +
  \left( {20881\over 144} - {17\over 12}\,\pi^2 - {35\over 2}\,\zeta_3
  \right) \, a(M_\tau^2)^2}
\no\\ &&\mbox{}   +
  \left( {13363099\over 5184} - {7927\over 144}\,\pi^2
  - {\pi^4\over 36} - {17318\over 27}\,\zeta_3
  + {715\over 12}\,\zeta_5 \right) \, a(M_\tau^2)^3 + \cdots
\no\\  && =
1 + 9.333\; a(M_\tau^2) + 109.989\; a(M_\tau^2)^2 +
1322.520\;  a(M_\tau^2)^3 + \cdots
\eeqn
The bold--guess estimate $d_4^L \sim d_3^L\, (d_3^L/d_2^L) \approx
4733$ would result in a huge $O(a^4)$ coefficient
$d_4^L + h^L_4 = 16\, 110$.

Since $\Delta^{L+T}$ has a larger weight on the total contribution to
$\delta^{(2)}_{us}$, the final combination of
the transverse and longitudinal pieces has a better behaviour\footnote{
The $O(a^2)$ correction agrees with the numerical result recently reported
in Ref.~\protect\cite{MA:98}, which is larger
than the value originally quoted in Ref.~\protect\cite{CK:93}.
This larger $O(a^2)$ correction has been also confirmed
by K. Chetyrkin and A. Kwiatkowski \protect\cite{CK:98}.
}:
\beqn\label{eq:Delta_res}
\lefteqn{\Delta[a(M_\tau^2)]  = 1 + {16\over 3} \; a(M_\tau^2) +
  \left( {10775\over 144} - {17\over 12}\,\pi^2 + {1\over 9}\,\zeta_3
  -{130\over 9} \,\zeta_5 \right) \, a(M_\tau^2)^2}\no \\
&&\mbox{}   +
  \left({3\over4} d_3^{L+T}+ {14050201\over 20736} - {5275\over 144}\,\pi^2
  - {\pi^4\over 144} - {47401\over 288}\,\zeta_3
  + {13195\over 432}\,\zeta_5 \right) \, a(M_\tau^2)^3\no \\
&&\mbox{}  + \cdots
\no\\ && =
1 + 5.333\; a(M_\tau^2) + 46.000\; a(M_\tau^2)^2 +
\left( 149.185 + {3\over 4} d^{L+T}_3 \right) a(M_\tau^2)^3 + \cdots\no \\
\eeqn
Nevertheless, the convergence of this perturbative series is very poor
for the range of the strong coupling relevant in $\tau$ decays,
$a(M_\tau^2) \sim 0.11$.
With $d^{L+T}_3 \sim d^{L+T}_2 \, ( d^{L+T}_2/d^{L+T}_1) \approx 317$,
 the $O(a^3)$
correction would be of the same size as the  $O(a)$ and $O(a^2)$
contributions.

\vspace{0.5cm}
\noindent {\bf 5. \,
Resummation of Running Effects along the Integration Contour}
\vspace{0.5cm}

At the moment, we can do very little about
the apparent growth of the $d^J_n$ coefficients, specially
for $J=L$. We clearly need a
deeper understanding of the perturbative
$D^J_{us}(s)\big|_{D=2}$ expansions.
However, we can try to control better the large contributions
contained in the $h_n^J$ factors.

The integration along the circle $x = e^{i\phi}$
gives rise to a long running of the quark mass and the  QCD coupling.
The expansion of $\ m^2(-\xi^2 M_\tau^2 x)\, a^n(-\xi^2 M_\tau^2 x)\ $
in powers of $a_\xi$ generates imaginary logarithms
$\log^n{(-x)} = i^n (\phi - \pi)^n$, which are large in some parts of
the integration range. The radius of convergence of such expansion is actually
quite small \cite{LDP:92a}.
However, there is no need to perform this ill--defined power expansion.

  Using in Eqs.~\eqn{eq:B_T} and \eqn{eq:B_L} the exact solution for
$m(-s)$ and $a(-s)$
obtained from the renormalization group equations,
the $B_J^{(n)}(a_\xi)$ integrals can be calculated to all orders in
$\alpha_s$, apart from the unknown $\beta_{n>4}$ and $\gamma_{n>4}$
contributions, which are likely to be small.
Thus, a more appropriate approach is to directly use the
expansions \eqn{eq:Delta}, in terms of the original
$\tilde{d}^J_n$ coefficients, and to fully keep the known four--loop
information on the functions $B_J^{(n)}(a_\xi)$.

Tables~\ref{tab:B_T} and \ref{tab:B_L} show the exact results
for $B_{L+T}^{(n)}(a)$ and $B_L^{(n)}(a)$ ($n=0, 1, 2, 3$) with $\xi=1$
obtained at different orders in the $\beta$ and $\gamma$ expansions,
together with the final values of $\Delta^J[a]$, for $a=0.1$  ($\xi = 1$).
For comparison the numbers coming from the truncated perturbative
expressions at $O(a^3)$ are also given.

\begin{table}[tbh]
\caption{Exact results for $B_{L+T}^{(n)}(a)$ \ ($n=0,1,2,3$) obtained at the
k--loop
($k=1,2,3,4$) approximation ($\beta_{j>k} = \gamma_{j>k} = 0$), together
with the final value of $\Delta^{L+T}[a] = \sum_{n=0}^2
d^{L+T}_n B_{L+T}^{(n)}(a)$,
for $a=0.1$ and $\xi=1$.
For comparison the numbers coming from the truncated
expressions at $O(a^3)$ are also given.}
\label{tab:B_T}
\vspace{0.2cm}
\centering
\begin{tabular}{|c|cccc|c|}
\hline
Loops & $B_{L+T}^{(0)}(a)$ & $B_{L+T}^{(1)}(a)$ &
$B_{L+T}^{(2)}(a)$ & $B_{L+T}^{(3)}(a)$
 & $\Delta^{L+T}[a]$
\\ \hline
1 & $0.890\, 32$ & $0.069\, 65$ & $0.004\, 52$ & $\phantom{-}0.000\, 186$ &
1.360
\\
2 & $0.817\, 19$ & $0.056\, 66$ & $0.002\, 78$ & $-0.000\, 008$ & 1.166
\\
3 & $0.791\, 43$ & $0.052\, 96$ & $0.002\, 36$ & $-0.000\, 048$ & 1.108
\\
4 & $0.782\, 37$ & $0.051\, 68$ & $0.002\, 22$ & $-0.000\, 060$ & 1.089
\\ \hline
$O(a^3)$
& $0.793\, 63$ & $0.064\, 73$ & $0.008\, 92$ & $\phantom{-}0.001\, 000$ & 1.405
\\ \hline
\end{tabular}
\end{table}


\begin{table}[tbh]
\caption{Exact results for $B_L^{(n)}(a)$ \ ($n=0,1,2,3$) obtained at the
k--loop
($k=1,2,3,4$) approximation ($\beta_{j>k} = \gamma_{j>k} = 0$), together
with the final value of $\Delta^L[a] = \sum_{n=0}^3 d^L_n B_L^{(n)}(a)$,
for $a=0.1$ and $\xi=1$.
For comparison the numbers coming from the truncated
expressions at $O(a^3)$ are also given.}
\label{tab:B_L}
\vspace{0.2cm}
\centering
\begin{tabular}{|c|cccc|c|}
\hline
Loops & $B_L^{(0)}(a)$ & $B_L^{(1)}(a)$ & $B_L^{(2)}(a)$ & $B_L^{(3)}(a)$
& $\Delta^L[a]$
\\ \hline
1 & $1.399\, 08$ & $0.184\, 73$ & $0.022\, 55$ & $0.002\, 588$ & 4.686
\\
2 & $1.540\, 13$ & $0.202\, 47$ & $0.024\, 21$ & $0.002\, 692$ & 5.052
\\
3 & $1.578\, 53$ & $0.206\, 17$ & $0.024\, 44$ & $0.002\, 690$ & 5.120
\\
4 & $1.589\, 10$ & $0.207\, 06$ & $0.024\, 46$ & $0.002\, 681$ & 5.133
\\ \hline
$O(a^3)$ &
$1.644\, 46$ & $0.218\, 94$ & $0.021\, 92$ & $0.001\, 000$ & 4.356
\\ \hline
\end{tabular}
\end{table}

These numerical results show a reasonable convergence of the $B_J^{(n)}(a)$
integrals, as higher--order
$\beta_k$ and $\gamma_k$ contributions are taken into account. Increasing
the number of loops one gets a small decrease (increase) of the
transverse (longitudinal) contribution. It is also clear that the truncated
$O(a^3)$ expressions overestimate (underestimate) $\Delta^{L+T}$ ($\Delta^L$).
Taking the full four--loop information
into account, we  get the following perturbative behaviour:
\bel{eq:Delta^T_beh}
\Delta^{L+T}[0.1] = 0.7824 + 0.2239 + 0.0823 - 0.0000601 \, d^{L+T}_3 + \cdots
\ee
\bel{eq:Delta^L_beh}
\Delta^L[0.1] = 1.5891 + 1.1733 + 1.1214 + 1.2489 + \cdots
\ee
The $L+T$ series converges very well. Owing to the negative running
contributions the $\Delta^{L+T}[a]$ series behaves better than the original
perturbative expansion of $D^{L+T}_{us}(s)\big|_{D=2}$. Unfortunately, the
longitudinal series is much more problematic. The bad perturbative behaviour
of $D^L_{us}(s)\big|_{D=2}$ gets reinforced by the running effects, giving
rise to a badly defined series.

The combined final expansion,
\bel{eq:Delta_beh}
\Delta[0.1] =  0.9840 + 0.4613 + 0.3421 +
\left( 0.3122 - 0.000045 \, d^{L+T}_3 \right) + \cdots
\ee
looks acceptable for the firsts terms because $\Delta^{L+T}$ is
weighted by a larger factor. In fact, this series behaves
better than the one in Eq.~(\ref{eq:Delta_res}), obtained with the
usual perturbative truncation of the contour integrals.
Nevertheless, after the third term the series appears to be dominated by
the longitudinal contribution,
and the bad perturbative behaviour becomes again manifest.

Using the full four--loop result we have certainly gained
in convergence for the $\Delta^{L+T}$ series
[compare the fourth term in the series \eqn{eq:Delta^T_exp} and
\eqn{eq:Delta^T_beh}  for $a=0.1$],
which is otherwise the one we don't know the $O(a^3)$ coefficient.
We can take advantage that the $O(a^3)$ correction to $\Delta[a]$
is almost completely
given by the known $\Delta^L$ contribution.
Using $d^{L+T}_3 \sim d^{L+T}_2 \, ( d^{L+T}_2/d^{L+T}_1) \approx 317$,
the fourth term in \eqn{eq:Delta_beh} becomes 0.298,
i.e. a 5\% reduction only.
Taking the size of the
$O(a^3)$ contribution to $\Delta^L$ as an educated estimate of
the perturbative uncertainty, we finally get
\bel{eq:result}
\Delta[0.1] = 2.1 \pm 0.3 \; .
\ee

\vspace{0.5cm}
\noindent {\bf 6. \, Renormalization--Scale Dependence}
\vspace{0.5cm}

The expansion (\ref{eq:Delta}) depends order by order on $\xi$ and this
dependence cancels out only when we sum the infinite series.
In practice, we only know a few first terms of the series
(three for $\Delta^{L+T}[a]$ and four for $\Delta^L[a]$);
so we should worry how much the predictions depend on our
previous choice $\xi=1$.
Obviously, $\xi$ should be close to one in order to avoid large logarithms;
but variations within a reasonable range, let us say from 0.75 to 2, should
not affect too much the final results. Smaller values of $\xi$ would put the
QCD coupling in the non-perturbative regime and are therefore  not acceptable.

\begin{figure}
\begin{center}
\leavevmode\epsfxsize=12cm\epsfbox{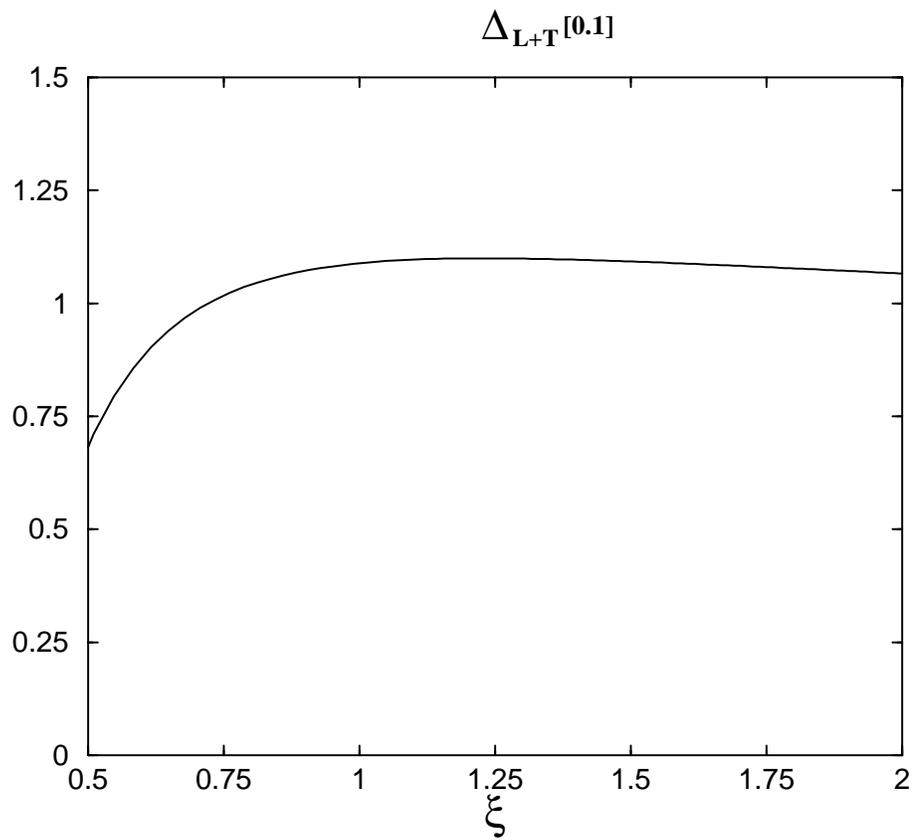}
\end{center}
\caption{\label{fig:DeltaLT}Variation of $\Delta_{L+T}[0.1]$
with the renormalization--scale factor $\xi$, to four loops.}
\end{figure}

\begin{figure}
\begin{center}
\leavevmode\epsfxsize=12cm\epsfbox{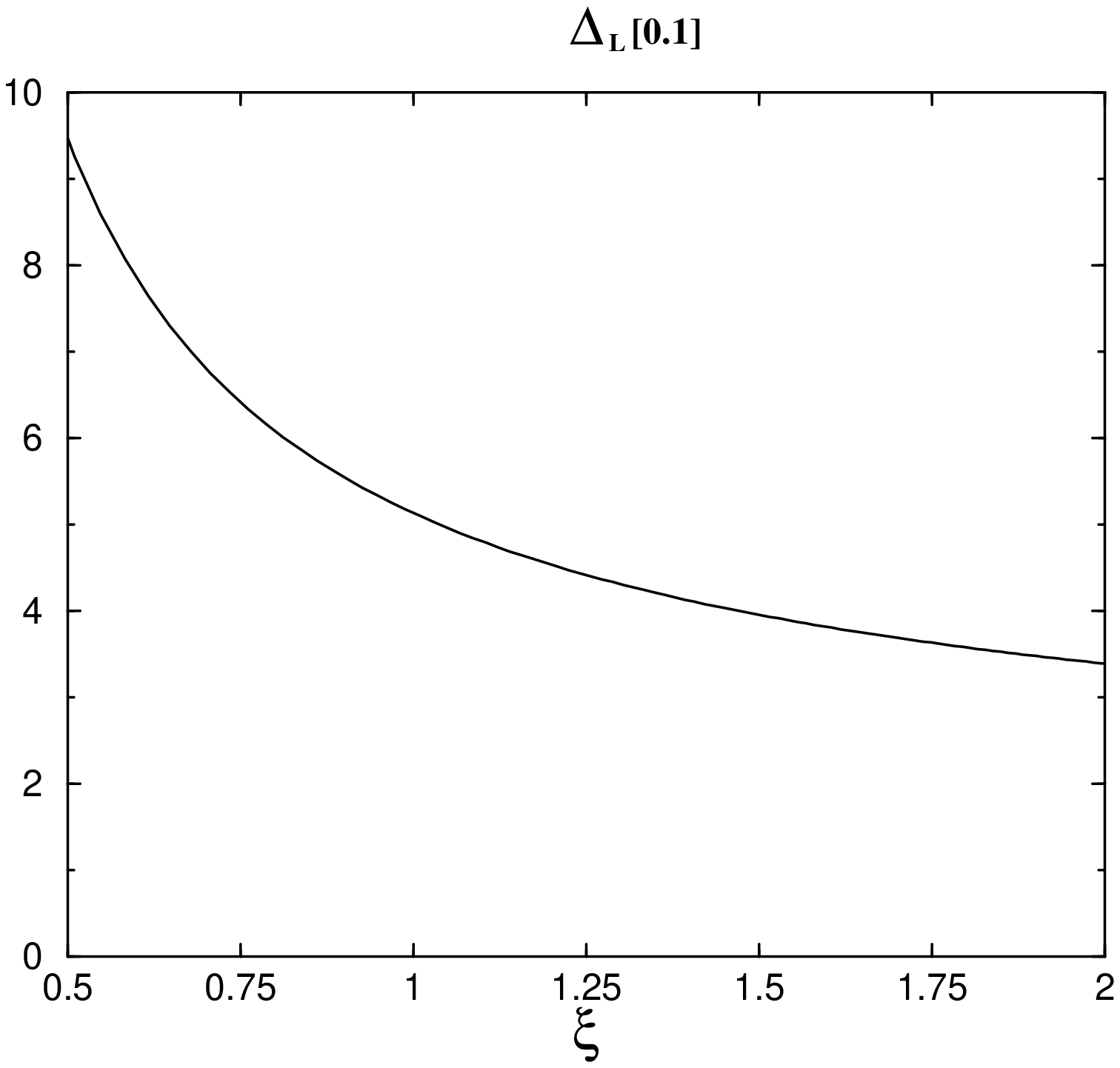}
\end{center}
\caption{\label{fig:DeltaL}Variation of $\Delta_{L}[0.1]$
with the renormalization--scale factor $\xi$, to four loops.}
\end{figure}

\begin{figure}
\begin{center}
\leavevmode\epsfxsize=12cm\epsfbox{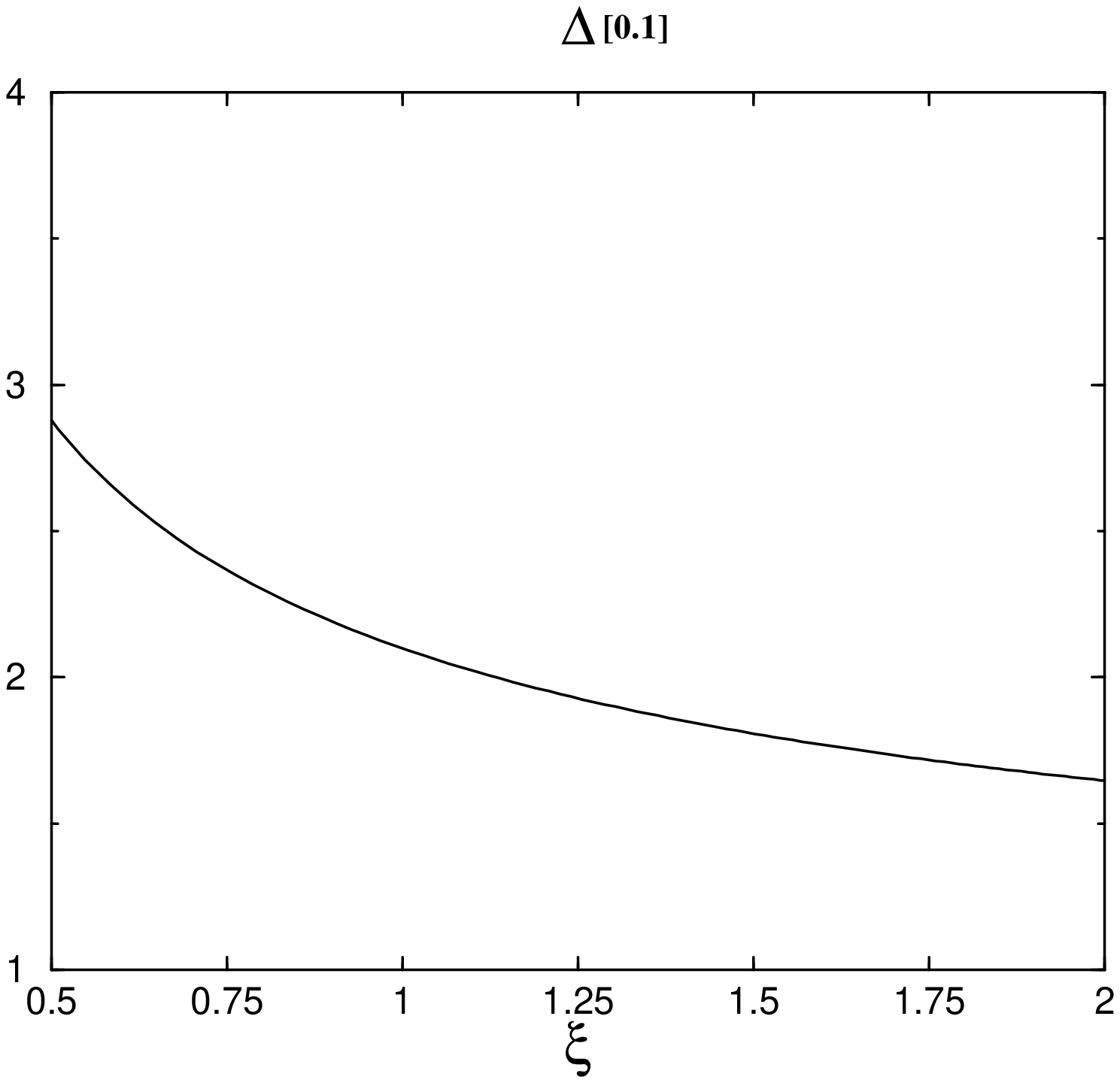}
\end{center}
\caption{\label{fig:Delta}Variation of $\Delta[0.1]$
with the renormalization--scale factor $\xi$,
to four loops.}
\end{figure}

Figures \ref{fig:DeltaLT}, \ref{fig:DeltaL} and \ref{fig:Delta}
show the sensitivity to the selection of renormalization scale
of the final predictions for  $\Delta^{L+T}$, $\Delta^L$,
and $\Delta$, respectively, for $a(M_\tau^2) = 0.1$.

The behaviour of $\Delta^{L+T}$ is quite good. The predicted value remains
very stable in the whole range $\xi\in [0.75,2]$, showing that the
perturbative series is very reliable. Below $\xi \sim 1/2$, the
perturbative expansion breaks down,
 as expected, because the coupling $a_\xi$ is already outside
the radius of convergence of the series.

The longitudinal series, on the other side, has a quite wild dependence on the
renormalization scale. Changing $\xi$ from 1 to 2, amounts to a reduction
of $\Delta^L$ of about 65\%. Thus, the theoretical uncertainty is very large
in this case.

The $\xi$ dependence of the complete expansion $\Delta$, reflects obviously
the behaviour of its two components. The larger weight of $\Delta^{L+T}$
keeps the result still acceptable, within the range of $\xi$ considered,
but the sizeable $\Delta^L$ contribution spoils the stability and generates
a monotonic decrease of the prediction for increasing values of $\xi$.
Taking this variation into account,
the theoretical error in Eq.~\eqn{eq:result} should be increased to about
0.6, i.e. a 30\% uncertainty in the final prediction.

\goodbreak
\vspace{0.5cm}
\noindent {\bf 6. \, Discussion}
\vspace{0.5cm}

The bad perturbative behaviour of the longitudinal contribution does not
allow to make an accurate determination of the strange quark mass from
$R_{\tau,S}$. Nevertheless, taking
\bel{eq:result_final}
\Delta[0.1] = 2.1 \pm 0.6 \; ,
\ee
$m_s(M_\tau^2)$ could be still obtained with a theoretical uncertainty
of about 15\%, which is not so bad.

Notice that it is the phase--space integration of the original correlation
functions the responsible for the different behaviour of the longitudinal
and transverse components. Therefore, the perturbative convergence could
probably be improved  through an appropriate use of weight factors
in Eqs.~\eqn{eq:spectral}  and \eqn{eq:circle}.
This requires an accurate measurement of the final hadrons mass distribution
in the $\tau$ decay, which so far has only been performed for the
dominant Cabibbo--allowed modes \cite{ALEPH:98}.
 The measurement of $R_{us}(s)$ could be feasible at the
forthcoming flavour factories, where a very good kaon identification is
foreseen.

~ From the theoretical point of view, the analysis of weighted moments of the
final hadrons mass distribution proceeds in a  completely analogous way
\cite{LDP:92b}.
A detailed study will be presented in a forthcoming publication.

\vspace{0.5cm}
\noindent {\bf Acknowledgements}
\vspace{0.5cm}

We would like to thank Kostja Chetyrkin for informing us about the
existence of a misprint in Ref.~\cite{CK:93}.
We have benefit from many discussions (and program
exchanges) with Matthias Jamin and Arcadi Santamar\'{\i}a, concerning
the numerical implementation of running effects at higher orders.
Useful discussions with Michel Davier, Andreas H\"ocker
 and Eduardo de Rafael are also
acknowledged. This work has been supported in part by
the European Union TMR Network  {\it EURODAPHNE}
---Contract No. ERBFMRX-CT98-0169 (DG 12 -- MIHT)---
and by CICYT, Spain, under Grants No. AEN-96/1672  and AEN-96/1718.

\newpage
\vspace{0.5cm}
\noindent {\bf Figure Captions}
\vspace{0.5cm}

\begin{itemize}
\item {\bf Figure 1.-}
Variation of $\Delta_{L+T}[0.1]$
with the renormalization--scale  factor $\xi$, to four loops.

\item {\bf Figure 2.-}
Variation of $\Delta_{L}[0.1]$
with the renormalization--scale factor $\xi$, to four loops.

\item {\bf Figure 3.-}
Variation of $\Delta[0.1]$
with the renormalization--scale factor $\xi$, to four loops.
\end{itemize}

\newpage
\pagestyle{empty}

\end{document}